\documentclass[twocolumn,aps,prl,showpacs]{revtex4}
\usepackage{epsf, epsfig, graphics,amsmath}

\begin{document}
\draft
\title{The internal states of active inclusions and the dynamics of an active membrane}
\author{Hsuan-Yi Chen}
\affiliation{Department of Physics and Center for Complex Systems,
National Central University, \\
Chungli, 32054 \\
Taiwan}

\date{\today}

\begin{abstract}
A theoretical model of a two-component fluid membrane containing
lipids and two-state active inclusions is presented.  Under strong
inclusion activities the membrane can be unstable due to
pump-driven undulation or aggregation of high curvature excited
inclusions. Depending on the structural details of the inclusions, 
the active conformation change of the inclusions can also induce 
finite size domains on the membrane.  
At long wavelengths a stable membrane has a height fluctuation spectrum 
similar to previous studies which neglected the inclusion internal states. 
For a tensionless membrane,
$\langle h({\bf q}_{\perp},
t)h(-{\bf q}_{\perp}, t)\rangle \sim T_{eff} q_{\perp}^{-4}$, where
$T_{eff}$ is an effective temperature which depends on the details of 
inclusion activities.
\end{abstract}

\maketitle
   A biological membranes is a  multi-component mixture of lipids, proteins,
carbohydrates, and other materials.~\cite{ref:Lodish_book} Since
the active inclusions (e.g., proteins) in the membranes participate
in physiological processes such as ion transport, signal
transduction, and cell locomotion, recent theoretical and
experimental studies have focused on the {\em collective effect}
of these active inclusions on the
membranes.~\cite{ref:Prost_00,ref:Prost_01,ref:Kumar_02} An
important conclusion of these studies is that, unlike equilibrium
membranes,~\cite{ref:Brochard_75} the height fluctuation of a 
tensionless active membrane in the long wavelength regime
depends on the inclusion activities.
Activity-induced instabilities are also predicted in previous
theoretical models.~\cite{ref:Prost_00,ref:Rao_01} On the other
hand, the fact that these active inclusions have more than one
internal conformation state has been discussed for
a single transmembrane protein~\cite{ref:Cantor_97} and for 
a single ion pump~\cite{ref:Park_02}.  Although the elastic
interaction between lipids and an active inclusion have
significant effect on the functions of a {\em single} inclusion,
the equally important question of how this interaction affects the
{\em collective} behavior of the lipid-inclusion system is seldom
addressed in previous studies of active
membranes.~\cite{ref:Mouritsen_98}

    In this Letter, I discuss the dynamics of a fluid membrane
containing lipids and identical two-state active inclusions. As
shown in Figure 1, inclusions in different conformation states
have different couplings to the local mean curvature of the membrane. 
Recent experiments on the structures of potassium
channels~\cite{ref:MacKinnon-1} and ${\rm K^+}$
channels~\cite{ref:Mackinnon-2} have provided structural evidences
of such inclusion conformation changes. A conformation change of
an inclusion can be induced by external stimuli (active
transition) or by the lateral pressure exerted by the surrounding
medium (passive transition). During a conformation change, an
inclusion also exerts forces to the lipids and the solvent. By
taking these effects into account, the main predictions of the
model are: (i) The height variance of a {\rm stable} tensionless
membrane in the long wavelength limit is $\langle |h({\bf
q}_{\perp},t)|^2 \rangle \sim T_{eff} /q_{\perp}^{4}$, where
$T_{eff}$ is an effective temperature which depends on the
inclusion activities. For a stable tense membrane, surface
tension dominates the long wavelength membrane height variance,
therefore  $\langle |h({\bf q}_{\perp},t)|^2 \rangle
\sim T/ q_{\perp}^{2}$, the same as that of an equilibrium
tense membrane. (ii) The system has two types of long
wavelength instabilities. They are
pump-driven undulation instability due to the pumping-induced
attraction between the inclusions, and the excitation-driven
undulation instability due to excitation-induced aggregation of
the excited inclusions in the case when the excited inclusions
have sufficiently higher curvature than the ground state
inclusions. (iii) Depending on the structural details of the 
inclusions, the system can have a
finite wavelength instability which leads to domains 
with a characteristic length depending on the active transition rates. 
At high active transition rates, the
activities can even drive the system to a homogeneous state.
Therefore it is possible to control the membrane morphology in
experiments by tuning the activities of the inclusions.

   In the Monge representation the shape of the membrane is described
by $h(x,y)$.   The number density of inclusions in state $\alpha $ 
($\alpha =1$ or $2$) is $\Phi_{\alpha}({\bf r}_{\perp},t)$ where ${\bf
r_{\perp}}=(x,y)$. It is useful to introduce
$\Phi _{\pm} = (\Phi _1 \pm \Phi _2)/2$ and $\Phi _{+}^{(0)}=\int
d^2r_{\perp}\Phi _{+}/\int d^2r$ . The coarse grained Hamiltonian
of the system $H = H_m + H_i + H_c$ includes the elastic
energy of the membrane, the direct interaction energy of the
inclusions, and the membrane-inclusion coupling.
To lowest order $H_m$ has the form
\begin{eqnarray}
  H_m
= \frac{1}{2} \int d ^2 r_{\perp} \left( \kappa
 (\nabla_{\perp} ^2 h)^2 + \gamma ({\bf \nabla}_{\perp}h)^2
                                  \right) ,
\end{eqnarray}
where $\kappa$ is the bending rigidity, $\gamma$ is the surface
tension of the membrane.  In this Letter I consider the regime
where the system is close to a phase separation in $\Phi _{-}$,
and $\Phi _{+}$ is non-critical. Therefore the interaction energy of
the inclusions in the harmonic theory is
~\cite{ref:H_i,ref:Lubensky_book}
\begin{eqnarray}
H_i = \int d^2 r_{\perp} \left\{ \right. && \frac{1}{2} (\xi
^2(\nabla_{\perp}\Phi _{-})^2 + r_2 \Phi _{-}^2 ) \nonumber \\
&&+ \frac{m}{2}(\Phi _{+} -\Phi _{+}^{(0)})^2 - \triangle
_{\epsilon}\Phi _{-} \left. \right\},
\end{eqnarray}
where $r_2 >0$ is small and $m \gg r_2$. $\triangle _{\epsilon}>
0$ is the excitation energy for an inclusion to change its conformation 
from state 1 to state 2. Thus ``state 1 (2)'' is the ``ground
(excited) state'' of an inclusion. The couplings between the
inclusion densities and the membrane curvature depend on the shapes 
of the inclusions, it can be expressed by 
\begin{eqnarray}
  H_c
&=& \int d^2 r_{\perp} \ \kappa (c_1 \Phi _1 + c_2 \Phi _2) \nabla
_{\perp}^2 h \nonumber \\
&=& \int d^2 r_{\perp} \ \kappa (c_+ \Phi _+ + c_{-} \Phi _{-})
                                         \nabla _{\perp}^2 h ,
\end{eqnarray}
where $c_{\pm} = c_1 \pm c_2$. 
The equilibrium properties of this model Hamiltonian is briefly
summarized in the following.  The stability criterion of the
membrane is $ c_2(q_{\perp})=
 \left(
 \kappa -\frac{\kappa ^2 c_{+}^2}{m}-\frac{\kappa ^2 c_{-}^2}{\xi q_{\perp}^2
 +r_2}
 \right) q_{\perp}^4
 + \gamma q_{\perp}^2 >0.$
For a tensionless membrane there is a long wavelength instability when
\begin{eqnarray}
\kappa _{eff} \equiv \kappa -\left( \frac{\kappa ^2 c_{+}^2}{m} +
 \frac{\kappa ^2 c_{-}^2}{r_2}\right) < 0,
\end{eqnarray}
where a phase separation in $\Phi _{-}$ occurs.
For a membrane under tension the system has a finite wavelength
instability when
\begin{eqnarray}
    \kappa _{eff}
<    -\left( \gamma \frac{\xi^2}{r_2}
           + 2 \sqrt{\kappa _{+} \left( \gamma \frac{\xi ^2}{r_2} \right) }
      \right),
\end{eqnarray}
where $\kappa _{+} \equiv \kappa - \kappa ^2 c_{+}^2/m$. In this
case, the long wavelength instability is cut off by the surface
tension at large length scales.
When the system is homogeneous, the equilibrium height fluctuation of
the membrane in the long wave limit is
$  <h(q_{\perp})h(-q_{\perp})>
 = \frac{k_B T}{\kappa _{eff} q_{\perp}^4}$ for a tensionless membrane
and
$  <h(q_{\perp})h(-q_{\perp})>
 = \frac{k_B T}{\gamma q_{\perp}^2}$ for a membrane under tension.

To discuss the dynamics, I introduce suitable variables $\phi
_{\pm}$.   When $c_2(q_{\perp})>0$, $\phi _{\pm}$ are chosen to be
the deviation from the equilibrium solution, i.e., $\phi _{+} =
\Phi _{+} -\Phi _{+}^{(0)}$ and $\phi _{-} = \Phi_{-} - \Phi
_{-}^{(0)}$, where $\Phi _{\pm}^{(0)}$ is the equilibrium value of
$\Phi _{\pm}$. When $c_2(q_{\perp})<0$, $\phi _{\pm}$ are chosen
to be $\phi_{+} = \Phi _{+} -\Phi _{+}^{(0)}$ and $\phi _{-} =
\Phi _{-}$ such that the effect of non-equilibrium activities on
the stability of the membrane can be discussed conveniently.   On
most experimental time scales the membrane does not exchange
inclusions with the solvent, therefore $\phi _+$ obeys the
conserved dynamics
\begin{eqnarray}
\frac{\partial \phi_+}{\partial t} = \lambda _+ \nabla _{\perp}
^2\frac{\partial H}{\partial \phi _+} +
\boldsymbol{\nabla}_{\perp} \cdot \boldsymbol{\zeta}_+,
\end{eqnarray}
where the vector thermal noise $\boldsymbol{\zeta}_+$ has zero
mean and variance 
$\langle 
\boldsymbol{\nabla}_{\perp}\cdot \boldsymbol{\zeta}_+ ({\bf r},t)
\boldsymbol{\nabla}_{\perp}\cdot \boldsymbol{\zeta}_+ ({\bf r'},t')
\rangle =
- 2 k_BT \lambda _{+} \nabla ^2 \delta ^3({\bf r}-{\bf r'})\delta (t-t')$
. In general $\phi _{-}$ satisfies an equation of the following form,
\begin{eqnarray}
\frac{\partial \phi _{-}}{\partial t} = (-\lambda _{0} + \lambda
_{2} \nabla _{\perp}^2) \frac{\delta H}{\delta \phi _{-}}-k_+ \phi
_+ - k_{-} \phi _{-} + \zeta _{-}.
\end{eqnarray}
The passive conformation change of the inclusions is taken into
account by $\lambda _0$ term,~\cite{ref:Ma_76} $\lambda _2$ term
represents the mutual diffusion of the inclusions, the
conformation changes due to external stimulations (active
transitions) are taken into account by $-k_+ \phi _+ - k_{-} \phi
_{-}$~\cite{ref:reaction}, the thermal noise $\zeta _{-}$ 
has zero mean and variance $2k_BT (\lambda _0 + \lambda
_2 q_{\perp}^2)$. The hydrodynamic flow, the permeation of solvent
through the membrane, and the force exerted on the membrane by the 
inclusions during active transitions all contribute
to the dynamics of the membrane, thus
\begin{eqnarray}
\frac{\partial h}{\partial t}= v_z - \lambda _p \left(
\frac{\delta H}{\delta h} +P_a^ek_e\phi _1 + P_a^r k_r \phi _2
\right) + \zeta _h ,
\end{eqnarray}
here $P_a^e$ ($P_a^r$) is the momentum transferred from an
inclusion during an  active
excitation (relaxation) process. The thermal noise $\zeta _h$
has zero mean and variance $2k_BT \lambda _p$. The solvent which
embeds the membrane satisfies the modified Stokes equation
\begin{eqnarray}
0 &=& - {\bf \nabla} p({\bf r},t)-\frac{\delta H}{\delta h}({\bf
r}_{\perp},t)\delta (z) {\bf \hat{z}} \nonumber \\&&+ P_a^e
[\delta (z-w_{1}^{u}) -\delta (z+w_{1}^{d}) ] k_e \phi _1 {\bf
\hat {z}} \nonumber \\ &&+P_a^r [\delta (z-w_{2}^{u}) -\delta
(z+w_{2}^{d}) ] k_r \phi _2 {\bf \hat {z}} \nonumber \\ &&+ \eta
\nabla ^2 {\bf v} + {\bf f}_{v}. \label{eq:v}
\end{eqnarray}
Here $p$ is the pressure, $\eta$ is the solvent viscosity, the
third and forth terms on the right hand side come from the active
transitions of the inclusions.  Although the total force acting on
the solvent due to an inclusion conformation change is zero,
the force distribution is not vanishing. To lowest order we
approximate the force distribution of an active inclusion
excitation (relaxation) event by a force
dipole.~\cite{ref:Prost_01} $w_1^u$, $w_1^d$, $w_2^u$, and $w_2^d$
are characteristic lengths associated with inclusion conformation
transition processes.  The vector thermal noise ${\bf f}_v$ has
zero mean and variance $\langle f_{vi}({\bf r},t)f_{vj}({\bf
r'},t') \rangle = 2k_BT \eta (-\delta _{ij}\nabla ^2 + \partial _i
\partial _j )\delta^3({\bf r}-{\bf r'})\delta (t-t')$.

Solving the Stokes equation~(\ref{eq:v}) and consider the regime
where the dynamics of $\phi _+$ is fast compared to that of $\phi
_{-}$,~\cite{note:fast},  the linearized equations of motion in
the Fourier space for $h$ and $\phi _{-}$ are obtained. Since the
contribution from permeation
in a typical system is negligible for lengths small compare
to $\mathcal{O}(1 {\rm cm})$~\cite{ref:Prost_01}, 
the equations for $h$ and $\phi _{-}$ becomes
\begin{eqnarray}
 \left( \begin{array}{c}
    \frac{\partial h(q_{\perp},t)}{\partial t} \\
    \frac{\partial \phi _{-}(q_{\perp}, t)}{\partial t}
       \end{array}
 \right)
=&\left( \begin{array}{ll}
  D_h(q_{\perp}) & D_{h\phi }(q_{\perp}) \\
  D_{\phi h}(q_{\perp}) & D_{\phi}(q_{\perp})
        \end{array}
 \right)
 \left( \begin{array}{c}
    h(q_{\perp},t) \\
    \phi _{-}(q_{\perp},t)
        \end{array}
 \right)& \nonumber \\
&+ \left( \begin{array}{c}
    f_h(q_{\perp},t) \\
    f_{\phi}(q_{\perp},t)
        \end{array}
 \right),&
\label{eq:eom}
\end{eqnarray}
where
\begin{eqnarray}
 && D_h(q_{\perp})
 = - \frac{1}{4\eta q_{\perp}}
     \left( \left(\kappa -\frac{\kappa ^2 c_{+} c_{+}^a}{m}\right) q_{\perp}^4
           + \gamma q_{\perp}^2 \right), \nonumber \\
 && D_{\phi}(q_{\perp})
 =-(\lambda _0 + \lambda _2 q_{\perp}^2)
                      (\xi^2 q_{\perp}^2 + r_{2})-k_{-}, \nonumber
                      \\
  &&D_{h\phi }(q_{\perp})
 = \frac{1}{4\eta q_{\perp}} \kappa c_{-}^a  q_{\perp}^2, \nonumber \\
{\rm and}&& \nonumber \\
 && D_{\phi h}(q_{\perp})
 = (\lambda _0 + \lambda _2 q_{\perp}^2) \kappa c_{-}q_{\perp}^2 -
   k_+ \frac{\kappa c_+}{m}q_{\perp}^2,
\end{eqnarray}
$\kappa c_{\pm}^a \equiv \kappa c_{\pm} + v_{\pm}$, and $v_{\pm} =
P_a^e k_e (-(w_1^u)^2+(w_1^d)^2)/2 \pm
P_a^rk_r(-(w_2^u)^2+(w_2^d)^2)/2$. It is convenient to think
$c_{\pm}^a$ as the ``renormalized'' inclusion-membrane coupling in
the presence of non-equilibrium activities.  Magnitudes of
$v_{\pm}$ depend on structural details and active transition rates of the 
inclusions. Dimensional analysis in previous
studies~\cite{ref:Prost_01,ref:Kumar_02} show that $v_{\pm} \sim
fl^2$, where $l\sim 5$nm is a typical length associated with an inclusion, 
and $f$ has the dimension of a force. At high activities $f \sim
10^{-12}$N~\cite{ref:Prost_01,ref:Kumar_02}, together with $\kappa
c_{\pm} \sim l k_BT$ gives $v_{\pm}/\kappa c_{\pm} \alt
\mathcal{O}(1)$. Therefore $\kappa c_{+}^a$ and $\kappa c_{+}$
($\kappa c_{-}^a$ and $\kappa c_{-}$) are of the same order of
magnitude, and usually have the same sign. The noises
$f_h(q_{\perp},t)$ and $f_{\phi}(q_{\perp},t)$ have zero mean and
variances $\langle f_h({\bf q}_{\perp},t) f_h({\bf
q'}_{\perp},t')\rangle =\frac{2k_BT}{4\eta q_{\perp}}\left(
1+v_{+}^2 q_{\perp}/4\eta \lambda _{+}m^2\right) \delta
(t-t')\delta^2({\bf q}_{\perp}+{\bf q'}_{\perp})$, and $\langle
f_{\phi}({\bf q}_{\perp},t)f_{\phi}({\bf q'}_{\perp},t')\rangle =
2k_BT\left( \lambda _0 + \lambda _2 q_{\perp}^2 + k_{+}^2/\lambda
_{+}m^2 q_{\perp}^2 \right) \delta (t-t') \delta^2({\bf
q}_{\perp}+{\bf q'}_{\perp})$. Notice that there are contributions
from the activities of the inclusions. They are no longer
equilibrium thermal noises.

The stability of the system is determined by the eigenvalues of
the $2 \times 2$ matrix in Eq.~(\ref{eq:eom}).  When both $D_h$
and $D_{\phi}$ are positive, the stability condition for a
tensionless membrane is
\begin{eqnarray}
&& \lambda _2 \frac{\xi ^2}{r_2} \kappa _{+}^a q_{\perp}^4 +\left(
   \lambda _0 \frac{\xi ^2}{r_2} \kappa _{+}^a +\lambda _2  \kappa _{eff}^a
 \right) q_{\perp}^2  \nonumber \\
&+&\left( \lambda _0 \kappa _{eff}^a +\frac{k_{+}}{r_2}
\frac{\kappa ^2 c_{+}c_{-}^a}{m} +\frac{k_{-}}{r_2} \kappa _{+}^a
\right)
>0.
\label{eq:stability}
\end{eqnarray}
where $\kappa _{+}^a= \kappa - \kappa ^2 c_{+} c_{+}^a/m$, $\kappa
_{eff}^a=  \kappa - \frac{\kappa ^2 c_{+}c_{+}^a}{m}
                             - \frac{\kappa ^2 c_{-}
                             c_{-}^a}{r_2}$.
It immediately follows that the system has a long wavelength
instability when $v_{+} \frac{\kappa c_{+}}{m}+ v_{-} \frac{\kappa
c_{-}}{r_2}$ is positive and sufficiently large such that the
$\kappa _{eff}^a$ is large and negative.
This instability is due to the pumping of the active inclusions.
As shown in Figure 2, when active pumping attracts more
inclusions, $\kappa _{eff}^a$ becomes negative. A similar
instability is discussed in a previous study which neglected
the internal states of the inclusions~\cite{ref:Prost_00}. Another
type of long wavelength instability occurs at $\kappa ^2 c_{+}
c_{-}^a<0$ and sufficiently large $k_{+} =k_e - k_r$.
Since $c_{-}^a$ is the renormalized $c_{-}$ in the non-equilibrium
situation, $\kappa ^2 c_{+} c_{-}^a \sim \kappa ^2 c_{+}c_{-}
<0$ can be interpreted intuitively as the case when ${c_2}^2
> {c_1}^2$, i.e., the excited state prefers higher membrane
curvature.  Thus, as Figure 3 shows, this instability is
due to the strong active excitations of the inclusions to their
higher curvature state.  

The system has a finite wavelength instability when the $q_{\perp}^4$ and
$q_{\perp}^0$ terms in Eq.~(\ref{eq:stability}) are positive and
\begin{eqnarray}
&&-\left(
   \lambda _0 \frac{\xi ^2}{r_2} \kappa _{+}^a
  +\lambda _2 \kappa _{eff}^a
\right) \nonumber \\
&>& 2\frac{\xi}{\sqrt {r_2}} \sqrt{\lambda _2 \kappa _{+}^a
 \left( \lambda _0 \kappa _{eff}^a
+\frac{k_{+}}{r_2} \frac{\kappa ^2 c_{+}c_{-}^a}{m}
+\frac{k_{-}}{r_2} \kappa _{+}^a
\right) }. \nonumber \\
\end{eqnarray}
The resulting steady state has an average domain size $R$ which 
can be estimated by the following scaling
analysis.~\cite{ref:Glotzer_95,ref:Lipowsky_01}  When there is no
budding, the growth of intramembrane domains in the absence of
inclusion activities corresponds to a two-dimensional phase
separation dynamics, i.e., $R(t)\sim t^{\alpha}$, $\alpha \approx
1/3$~\cite{ref:Seul_94}. This growth eventually saturates due to the active
transitions. Since the time scales associated with active
transitions are $k_{\pm}$, the steady state domain size $R$ should
obey $R\sim k_M^{-\alpha}$, where $k_M = k_{+(-)}$ when
$k_{+}\kappa ^2 c_{+}c_{-}^a/m$ is greater (smaller) than
$k_{-}\kappa _{+}^a$. Notice that this instability is suppressed at 
large $k_{-} = k_e+ k_r$. This is because when the inclusions
change their conformations at extremely high rates, the membrane
feels the time-averaged inclusion conformation, therefore there is
no finite size domains in the steady state. 
When $\kappa ^2 c_{+}c_{-}^a >0$, an excited state inclusion
has smaller curvature than a ground state inclusion,sufficiently 
large $k_{+}$ stabilizes the membrane because strong active transitions 
excite the inclusions to the smaller curvature excited state. 

When the membrane is stable, its height
fluctuations in the long wavelength limit is expressed by
$\frac{k_BT_{eff}}{\kappa _{eff} q_{\perp}^4}$,
where
\begin{eqnarray}
  T_{eff}
= \frac{T \kappa _{eff}}{
  \kappa -\frac{\kappa ^2 c_{+}c_{+}^a}{m}
- \frac{\kappa ^2 c_{-}c_{-}^a - k_{+} \kappa ^2c_{+}c_{-}^a/\lambda _0 m}
       {r_2 - k_{-}/\lambda _0} }.
\label{eq:Teff1}
\end{eqnarray}
If the passive conformation changes of the inclusions are extremely rare
such that $\lambda _0 \ll \lambda _2 q_{\perp}^2$ in all experimental length
scales, $T_{eff}$ has the following form,
\begin{eqnarray}
 T_{eff}
= \frac{T \kappa _{eff}}{
  \kappa
 -\frac{\kappa ^2 c_{+}c_{+}^a}{m}-\frac{k_{+}}{k_{-}}
     \frac{\kappa ^2c_{+}c_{-}^a}{m} }.
\label{eq:Teff2}
\end{eqnarray}
Eqs.~(\ref{eq:Teff1})(\ref{eq:Teff2}) are similar to previous studies on the 
active membranes which neglected
the internal states of the inclusions.~\cite{ref:Kumar_02}
For a membrane under tension the $\frac{k_BT_{eff}}{\kappa q_{\perp}^4}$
behavior is cut off at long wavelengths.  The membrane height fluctuation
at large lengths in this case is simply $k_BT/\gamma q_{\perp}^2$ where $T$
is the temperature of the solvent.

In summary, a theory of active membranes with two-state active inclusions is
discussed. Although current analysis is restricted to a
relatively small region of the parameter space, the rich dynamical 
behaviors shown in the analysis are believed to exist in more general 
situations. These include
instability induced by active pumping, instability induced by
inclusion-excitation, finite size domains controlled by activities,
and activity-restored homogeneous morphology. In the stable case,
the membrane height fluctuation is similar to several
previous theoretical and experimental studies.  The analysis of this 
model in a broader parameter range and the effects of nonlinearity 
will be included in a future work.~\cite{ref:future} It
is my hope that this model might be of relevance to active
conformational transitions in biological membranes, and
experimental work on artificial membranes might be designed to
confirm the main ideas of this model.

I would like to thank Prof. David Jasnow for his encouragement and
stimulating discussions. I also thank Prof. Peilong Chen for
interesting discussions.  This work is supported by National
Science Council of the Republic of China under grant
number NSC-91-2112-M-008-052.

\vskip 1in

\noindent
{\large Figures}
\begin{figure}[h]
\epsfxsize= 3 in
\epsfbox{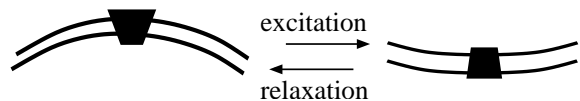}
\caption{Schematics of an
inclusion in different conformation states}
\end{figure}

\begin{figure}[h]
\epsfxsize= 3 in
\epsfbox{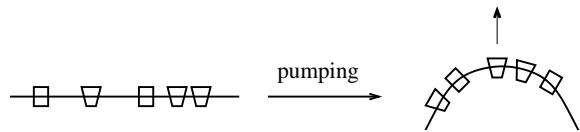}
\caption{Large negative $\kappa _{eff}^a$ instability: pumping-induced 
attraction between inclusions induces inclusion aggregation.
}
\end{figure}

\begin{figure}[h]
\epsfxsize= 3 in
\epsfbox{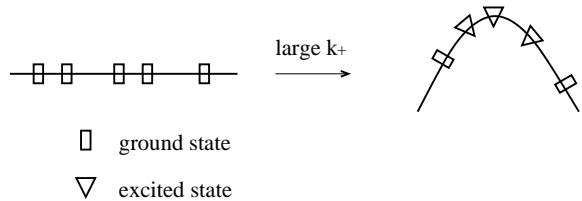}
\caption{The presence of many high-curvature excited state inclusions
induces a long wavelength instability.}
\end{figure}

\begin{references}
\bibitem{ref:H_i}  In general there can a term $\Phi _{+} \Phi
_{-}$ due to different density susceptibilities of  $\Phi _1$ and
$\Phi _2$.  The effect of this density susceptibility difference
will be discussed in a future work.~\cite{ref:future}
\bibitem{ref:reaction} The active conformation changes of the inclusions
contribute to the dynamics of $\phi _1$ ($\phi _2$) through terms
of the form $-k_e \phi _1 + k_r \phi _2$ ($+k_e \phi _1 - k_r \phi
_2$). Thus $k_{\pm} = k_e {\mp}k_r$. The active transition rates 
$k_{e}$, $k_r$ depend on the local strength of external stimuli, 
which is related to the
concentration of specific ligands. Since the diffusion of the
ligands in the solvent is fast compare to the in-plane dynamics
of the inclusions, in this Letter I treat both $k_e$ and
$k_r$ as constants.
\bibitem{note:fast}
The dynamics of $\phi _{-}$ is slow compare to that of $\phi _{+}$
when $(\lambda _0 + \lambda _2 l_{max}^{-2})r_2 + k_{-} \ll \lambda
_{+}m l_{max}^{-2}$, where $l_{max}$ is the longest wavelength in the 
experiments.  Since $\lambda _{+}/\lambda _{2} \sim \mathcal{O}(1)$,
this condition is satisfied when 
$\sqrt{\lambda _{+}m/(\lambda _0 r_2 + k_{-})} \gg l_{max}$,
\bibitem{note:Mouritsen} A similar effect has been discussed in
Ref.~\cite{ref:Mouritsen_98}, where two-state active inclusions with
different hydrophobic length were considered.
\end{references}
\end{document}